\documentstyle[12pt]{article}

\catcode`\@=11
\def\marginnote#1{}
\newcount\hour
\newcount\minute
\newtoks\amorpm
\hour=\time\divide\hour by60
\minute=\time{\multiply\hour by60 \global\advance\minute by-\hour}
\edef\standardtime{{\ifnum\hour<12 \global\amorpm={am}%
        \else\global\amorpm={pm}\advance\hour by-12 \fi
        \ifnum\hour=0 \hour=12 \fi
        \number\hour:\ifnum\minute<10 0\fi\number\minute\the\amorpm}}
\edef\militarytime{\number\hour:\ifnum\minute<10 0\fi\number\minute}
\def\draftlabel#1{{\@bsphack\if@filesw {\let\thepage\relax
   \xdef\@gtempa{\write\@auxout{\string
      \newlabel{#1}{{\@currentlabel}{\thepage}}}}}\@gtempa
   \if@nobreak \ifvmode\nobreak\fi\fi\fi\@esphack}
        \gdef\@eqnlabel{#1}}
\def\@eqnlabel{}
\def\@vacuum{}
\def\draftmarginnote#1{\marginpar{\raggedright\scriptsize\tt#1}}
\def\draft{\oddsidemargin -.5truein
        \def\@oddfoot{\sl preliminary draft \hfil
        \rm\thepage\hfil\sl\today\quad\militarytime}
        \let\@evenfoot\@oddfoot \overfullrule 3pt
        \let\label=\draftlabel
        \let\marginnote=\draftmarginnote
   \def\@eqnnum{(\theequation)\rlap{\kern\marginparsep\tt\@eqnlabel}%
\global\let\@eqnlabel\@vacuum}  }

\def\preprint{\twocolumn\sloppy\flushbottom\parindent 1em
        \leftmargini 2em\leftmarginv .5em\leftmarginvi .5em
        \oddsidemargin -.5in    \evensidemargin -.5in
        \columnsep 15mm \footheight 0pt
        \textwidth 250mmin      \topmargin  -.4in
        \headheight 12pt \topskip .4in
        \textheight 175mm
        \footskip 0pt
        \def\@oddhead{\thepage\hfil\addtocounter{page}{1}\thepage}
        \let\@evenhead\@oddhead \def\@oddfoot{} \def\@evenfoot{} }

\def\titlepage{\@restonecolfalse\if@twocolumn\@restonecoltrue\onecolumn
     \else \newpage \fi \thispagestyle{empty}\c@page\z@ 
        \def\thefootnote{\fnsymbol{footnote}} }

\def\endtitlepage{\if@restonecol\twocolumn \else  \fi
        \def\thefootnote{\arabic{footnote}}
        \setcounter{footnote}{0}}  

\catcode`@=12
\relax
\def\beq{\begin{equation}}
\def\eeq{\end{equation}}
\def\Im{\mathop{\rm Im}}
\def\NP#1#2#3{Nucl. Phys. \underline{#1} (19#2) #3}

\def\PL#1#2#3{Phys. Lett. \underline{#1} (19#2) #3}
\def\PR#1#2#3{Phys. Rev. \underline{#1} (19#2) #3}

\def\Re{\mathop{\rm Re}}

\def\crbig{\\\noalign{\vspace {3mm}}}
\relax

\begin{document}
\topmargin-2.4cm
\renewcommand{\theequation}{\thesection.\arabic{equation}}
\begin{titlepage}
\begin{flushright}
NEIP--00--007 \\
hep--th/0003168 \\
March 2000
\end{flushright}
\vspace{2.4cm}

\begin{center}
{\Large\bf M-Theory N=1 Effective Supergravity}
\vskip .1in
{\Large\bf in Four Dimensions$^\star$}
\vskip .8in
{\bf J.-P. Derendinger and R. Sauser}$^{\dagger}$
\vskip .1in
{\it Institute of Physics,
University of Neuch\^atel \\
CH--2000 Neuch\^atel, Switzerland}
\end{center}

\vspace{1.9cm}

\begin{center}
{\bf Abstract}
\end{center}
\begin{quote}
We present a general `off-shell' description of the effective 
$N=1$ supergravity describing the low-energy limit of M-theory 
compactified on $\hbox{(Calabi-Yau)} \times S^1/{\bf Z}_2$. 
In our formulation, the M-theory Bianchi identities are imposed by
the equations of motion of four-dimensional supermultiplets. 
Modifications of these identities
(resulting for instance from contributions localized at orbifold 
singularities or non-perturbative sources like five-branes) can then 
easily be implemented.
\end{quote}

\vfill
\begin{flushleft}
\rule{8.1cm}{0.2mm}\\
$^{\star}$
{\small Talk given by R. Sauser at the TMR meeting 
on Quantum Aspects of Gauge Theories, Supersymmetry and Unification, 
Paris, 1--7 September 1999.} \\
$^{\dagger}$ {\small\tt jean-pierre.derendinger, roger.sauser@iph.unine.ch}
\end{flushleft}

\end{titlepage}
\setcounter{footnote}{0}
\setcounter{page}{0}
\setlength{\baselineskip}{.7cm}
\newpage

\section{Introduction and conclusions}\label{secintroandconclusions}
\setcounter{equation}{0}

M-theory compactified on $O_7 \equiv X_6 \times S^1/{\bf Z}_2$, where 
$X_6$ is a Calabi-Yau (CY) three-fold, leads to a four-dimensional 
theory with $N=1$ local supersymmetry.
In the low-energy limit, M-theory information can be organized as an 
expansion in powers of the eleven-dimensional gravitational constant 
$\kappa_{11}$ \cite{HW1, HW2}. The lowest order $\kappa^{-2}_{11}$ is 
eleven-dimensional supergravity \cite{CJS}. In a compactification on 
$S^1/{\bf Z}_2$ only, the next orders are known to include orbifold 
plane contributions  as well as gauge 
and gravitational anomaly-cancelling terms \cite{HW1, W, HW2}.
Similarly, the effective four-dimensional supergravity can be
formulated as an expansion in the four-dimensional gravitational 
constant $\kappa$, even if string theory rather suggests to use 
the dilaton as expansion parameter. 
The lowest order $\kappa^{-2}$ is the $S^1/{\bf Z}_2$ truncation 
of eleven-dimensional supergravity on a CY three-fold. 
The next order includes super-Yang-Mills (SYM) and charged matter kinetic 
and superpotential contributions. Sigma-model anomaly-cancelling terms 
modifying in particular the gauge thresholds are then also involved. 
These first corrections to the low-energy limit of M-theory 
compactifications on $O_7$ are identical to those obtained from 
heterotic compactifications on CY. The literature gives a detailed 
description of these results, with particular attention paid to the 
`strong-coupling heterotic limit' in which the size of the CY space 
is smaller than the orbifold length, supersymmetry breaking by 
gaugino condensation and non-standard embeddings 
\cite{ref1}--\cite{nonstandard}.

In this note, we give the structure of the four-dimensional $N=1$ 
wilsonnian effective supergravity describing the universal massless 
sector of M-theory compactified on $O_7$.
We begin by writing the theory corresponding to the reduction 
of the bulk eleven-dimensional supergravity directly in terms of 
four-dimensional `M-theory supermultiplets'. The supersymmetrized Bianchi 
identities for the components of the M-theory tensor field strength are 
promoted to equations of motion using `Lagrange multiplets'. 
Within this `off-shell' approach, we can then introduce `source multiplets' 
to take into account the contributions of the $S^1/{\bf Z}_2$ planes 
which appear as modifications of the Bianchi identities. This formulation 
is also particularly appropriate for the inclusion of non-perturbative 
states (M-theory five-branes, condensates, etc.).

The material presented here is detailed in ref. \cite{DS1} and 
a forthcoming publication \cite{DS2} will contain a direct application 
of our approach (the coupling of five-brane moduli to the background).

\section{The bulk Lagrangian}\label{secsugra}
\setcounter{equation}{0}

In this section, we establish our basic procedure by considering the 
well-known `bulk dynamics', which follows from $O_7$ compactification 
of eleven-dimensional supergravity. The resulting Lagrangian is the 
lowest order in the $\kappa$-expansion and describes Kaluza-Klein (KK) 
massless modes of eleven-dimensional supergravity. 

We will precisely describe two aspects which may be of importance in 
M-theory compactifications.
Firstly, we will introduce chiral, linear or vector supermultiplets with 
constraints in order to obtain a supersymmetric version of the Bianchi 
identities satisfied by antisymmetric tensors.
Secondly, we will use superconformal supergravity in which we can keep 
open the choice of gravity frame.

\subsection{Superconformal formalism} 

We use the superconformal formulation of $N=1$ supergravity with a 
chiral compensating multiplet $S_0$ (with conformal and chiral 
weights $w=1$ and $n=1$) to generate Poincar\'e theories by gauge 
fixing. In this formalism, a change of frame corresponds to a 
different Poincar\'e gauge condition applied on the modulus of 
the scalar compensator $z_0$, which fixes dilatation symmetry. 
Up to terms with more than two derivatives and up to terms which 
would contribute to kinetic terms in a fermionic background 
only \cite{CFGVP1, DFKZ1}, the most general supergravity Lagrangian 
reads\footnote{
Except otherwise mentioned, our notations for superconformal 
expressions are as in refs. \cite{KU}, from where the original 
literature can also be traced back. The appendix of ref. \cite{DS1} 
displays the conventions we follow through this note.}
\begin{equation}\label{Phidef1}
{\cal L} = \left[ S_0\overline S_0 \Phi \right]_D 
+ \left[ S_0^3 W \right]_F
+{1\over4}\left[f_{ab}{\cal W}^a{\cal W}^b\right]_F.
\end{equation}
The symbols $[\ldots]_D$ and $[\ldots]_F$ denote the invariant 
$D$- and $F$-density formulas given by (all fermion contributions 
are omitted)
\begin{equation}
[{\cal V}]_{D} = e(d+{1 \over 3} cR) ~~\hbox{ and }~~
[{\cal S}]_{F} = e(f+\overline f),
\end{equation}
where $\cal V$ is a vector multiplet with components 
$(c,\chi,m,n,b_{\mu},\lambda,d)$ and $\cal S$ a chiral multiplet 
with components $(z,\psi,f)$. The real vector multiplet $\Phi$ 
(zero weights) is a function (in the sense of tensor calculus) of 
the multiplets present in the theory, including in general the 
compensating multiplet. The holomorphic function $W$ of the chiral 
multiplets is the superpotential. The chiral multiplet $\cal W$ is 
the gauge field strength for the gauge multiplets and $f_{ab}$ is 
the holomorphic gauge kinetic function of the chiral multiplets.
Besides $S_0$ and $\cal W$, we will use chiral multiplets with zero 
weights and neither $W$ nor $f_{ab}$ will depend on the compensator. 

Using a $U(1)$/K\"ahler gauge fixing the supergravity Lagrangian 
(\ref{Phidef1}) can also take the form
\begin{equation}\label{Phidef2}
{\cal L} = \left[ S_0\overline S_0 \Phi \right]_D 
+ c \left[ S_0^3 \right]_F 
+{1\over4}\left[f_{ab}{\cal W}^a{\cal W}^b\right]_F,
\end{equation}
with an arbitrary constant $c$ as superpotential and two arbitrary 
functions $\Phi$ and $f_{ab}$.

\subsection{Supermultiplets with constraints}

The Lagrangian of eleven-dimensional supergravity can be written 
as \cite{CJS}
\begin{equation}\label{11dsugraCJS}
\begin{array}{rcl}
e^{-1}{\cal L}_{\rm CJS} &=&  
-{1\over2\kappa_{11}^2}R 
-{1\over 4\kappa_{11}^2}{1\over 4!}\,
G_{M_1M_2M_3M_4}G^{M_1M_2M_3M_4}
\crbig
&&-{1\over 12\kappa_{11}^2}
\,{1\over4!4!3!}\,e^{-1}\epsilon^{M_1\ldots M_{11}}
G_{M_1M_2M_3M_4}
G_{M_5M_6M_7M_8}C_{M_9M_{10}M_{11}}
\crbig
&&+\,{\rm fermionic\,\,terms}.
\end{array}
\end{equation}
Omitting all fields related to the detailed geometry of the CY
manifold, the particle content of the four-dimensional theory is the 
$N=1$ supergravity multiplet, with metric tensor $g_{\mu\nu}$, and 
matter multiplets including on-shell four bosons and four fermions. 
Two bosons are scalars and correspond to the dilaton and the 
`universal modulus' of the CY space, the massless volume mode. 
Two bosons are KK modes of the field strength $G$, with Bianchi 
identity $dG=0$. Explicitly, these two last fields and their Bianchi 
identities read\footnote{
In our notations, $x^4$ is the orbifold coordinate.
}
\begin{equation}\label{f4modes1}
\begin{array}{lcl}
G_{\mu\nu\rho 4}, &\qquad& 
\partial_{[\mu}G_{\nu\rho\sigma 4]} = 0, \crbig
G_{\mu j \overline k 4} = i\, T_\mu\, \delta_{j\overline k}, && 
\partial_{[\mu} T_{\nu]} = 0. 
\end{array}
\end{equation}
It will prove useful to identify these fields with the vector
components of two real vector multiplets $V$ ($w=2$, $n=0$) and $V_T$ 
($w=n=0$), and to impose the Bianchi identities as field equations 
using a chiral multiplet $S$ ($w=n=0$) and a real linear multiplet 
$L_T$ ($w=2$, $n=0$) as Lagrange multipliers. The bulk supergravity 
Lagrangian takes then the form
\begin{equation}\label{LconfCY}
{\cal L}_{\rm B} = 
\left[
- (S_0\overline S_0V_T)^{3/2}(2V)^{-1/2}
- (S+\overline S)V + L_TV_T 
\right]_D.
\end{equation}
The various superconformal multiplets appearing in this Lagrangian 
have the following components expressions\footnote{
We only explicitly consider the bosonic sector of the theory and omit 
all fermions in the $N=1$ supermultiplets. We gauge-fix the 
superconformal symmetries not contained in $N=1$ Poincar\'e 
supersymmetry, except dilatation symmetry. Notice also that 
our component expansion of vector multiplets differs in its 
highest component from refs. \cite{KU}.
}
\begin{equation}\label{comp1}
\begin{array}{rcl}
V &=& (C,0,H,K,v_\mu,0, d-\Box C-{1\over3}CR), \crbig 
V_T &=& (C_T,0,H_T,K_T,T_\mu,0, d_T-\Box C_T), \crbig 
S &=& (s,0,-f,if,i\partial_\mu s,0, 0), \crbig 
L_T &=& (\ell_T,0,0,0,t_\mu,0, -\Box\ell_T-{1\over3}\ell_T R), 
\crbig
S_0 &=& (z_0,0,-f_0,if_0,iD^c_\mu z_0,0, 0). 
\end{array}
\end{equation}
The role of the Lagrange multipliers $S$ and $L_T$ follows from 
the two relations
\begin{equation}\label{SLT}
\begin{array}{rcl}
e^{-1}[(S+\overline S)V]_D &=&
-2\Im s \,\partial^\mu v_\mu + 2d \Re s 
-f(H-iK) -\overline f(H+iK)
\crbig
&&+\,{\rm derivative}, \crbig
e^{-1}[L_TV_T]_D &=& \ell_T(d_T-\Box C_T) 
-{e\over2}\epsilon_{\mu\nu\rho\sigma}(\partial^\mu T^\nu)\, 
t^{\rho\sigma} + {\rm derivative}.
\end{array}
\end{equation}
In the last equality, we have used the constraint imposed to the 
linear multiplet $L_T$, $\partial^\mu t_\mu=0$, to write 
$t_\mu={e\over2} \epsilon_{\mu\nu\rho\sigma}\partial^\nu  t^{\rho\sigma}$. 
Solving for the components of $S$ leads to 
$\partial^\mu v_\mu=d=H=K=0$, 
and $V$ is a linear multiplet $L$ ($w=2$, $n=0$). 
Solving for the components of $L_T$ leads to $d_T-\Box C_T= 
\partial_{[\mu} T_{\nu]}=0$, and $V_T$ can be written as 
$T+\overline T$, with a chiral weightless multiplet $T$\footnote{
With components: $C_T=2\Re T$, $T_\mu=-2\partial_\mu \Im T$, 
$H_T=-2\Re f_T$, $K_T=-2\Im f_T$.}.
Since one can always write $v_\mu={e\over6}\epsilon_{\mu\nu\rho\sigma} 
v^{\nu\rho\sigma}$, we have generated with $\Im s$ and $t_{\mu\nu}$
the Bianchi identities 
$\partial_{[\mu}v_{\nu\rho\sigma]} = \partial_{[\mu} T_{\nu]} = 0$. 
A modification of these Bianchi 
identities, as induced by $S^1/{\bf Z}_2$ compactification or by 
five-brane couplings will then be phrased as a modification of the 
supermultiplets appearing multiplied by $S+\overline S$ or $L_T$ in 
Eqs. (\ref{SLT}).

The structure of the Lagrangian (\ref{LconfCY}) reflects the familiar 
duality relating scalars and antisymmetric tensors or, for
superfields, chiral and linear multiplets. 

Solving in Eq. (\ref{LconfCY}) for the Lagrange multipliers $S$ and 
$L_T$ leads to the `standard form' of the bulk four-dimensional 
Lagrangian \cite{CFV, DQQ}
\begin{equation}\label{4dLCJS}
{\cal L}_{\rm B,l} = -\left[ 
\bigl(S_0\overline S_0 \, e^{-\hat K/3}\bigr)^{3/2}
(2L)^{-1/2}\right]_D,
\end{equation}
with the K\"ahler potential $\hat K=-3\log(T+\overline T)$ for the 
volume modulus $T$. We will see again below that this standard form 
is naturally obtained by direct reduction of the Cremmer, Julia and 
Scherk version of eleven-dimensional supergravity on $O_7$. Clearly, 
theory (\ref{4dLCJS}) is also the CY truncation of ten-dimensional 
$N=1$ pure supergravity \cite{CFV}.

Solving for $V$ and $L_T$ in Eq. (\ref{LconfCY}) leads to the familiar 
chiral form \cite{W4dsugra}
\begin{equation}\label{bulkchiral}
{\cal L}_{\rm B,c} = 
-{3\over2}\left[S_0\overline S_0 \, e^{-K/3}\right]_D,
\end{equation}
with $K=-\log(S+\overline S)+\hat K$.

\subsubsection{Choice of Poincar\'e frame}\label{secEinstein}

According to the component expression for the $D$-density and 
the tensor calculus of superconformal multiplets \cite{KU}, the 
Einstein term included in the bulk Lagrangian (\ref{LconfCY}) is 
\cite{FGKVP, DQQ}
\begin{equation}\label{Eins2}
{\cal L}_{\rm E} = -{1\over2}eR\left[ 
\left( z_0\overline z_0 C_T \right)^{3/2}\left(2C\right)^{-1/2} 
\right] .
\end{equation}
As they should, the terms introduced to impose Bianchi identities do 
not contribute. We then select the Einstein frame, in which 
the gravitational Lagrangian is $-{1\over2\kappa^2}eR$, by the 
dilatation gauge condition 
\begin{equation}\label{Eins3}
\kappa^{-2} = (z_0\overline z_0 C_T)^{3/2}(2C)^{-1/2}.
\end{equation}
It will be convenient to introduce the (composite) real vector multiplet
\begin{equation}\label{Upsilonis}
\Upsilon = (S_0\overline S_0 V_T)^{3/2} (2V)^{-1/2},
\end{equation}
with conformal weight two. In the Poincar\'e theory and in the
Einstein frame, its lowest component is equal to $\kappa^{-2}$.

\subsubsection{Identification of the components}

Choosing the Einstein frame, $\Upsilon=\kappa^{-2}$, and solving for 
the components of $S$ and $L_T$, the complete bosonic expansion of 
the four-dimensional supergravity (\ref{LconfCY}) is
\begin{equation}\label{bulkcompfinal}
\begin{array}{rcl}
e^{-1}{\cal L}_{\rm B} &=& 
- {1 \over {2\kappa^2}}  R
- {1 \over {4\kappa^2}} C^{-2} 
	[ (\partial_\mu C)(\partial^\mu C) - v_\mu v^\mu ] \crbig
&&- {3 \over {4\kappa^2}} C_T^{-2} 
	[ (\partial_\mu C_T)(\partial^\mu C_T) + T_\mu T^\mu ],
\end{array}
\end{equation}
with $v_\mu = {e \over 2}\epsilon_{\mu\nu\rho\sigma}\partial^\nu 
b^{\rho\sigma}$ since $V$ is a linear multiplet, $C_T = 2\Re T$ 
and $T_\mu = -2\partial_\mu \Im T$ since $V_T = T+\overline T$.

This Lagrangian is to be compared with the one we obtain from the 
reduction of eleven-dimensional supergravity
(\ref{11dsugraCJS}). The ${\bf Z}_2$ orbifold projection eliminates
all states which are odd under $x^4\rightarrow -x^4$, and the
reduction of the eleven-dimensional space-time metric is
\begin{equation}\label{D11metric}
g_{MN} = \pmatrix{
e^{-\gamma} e^{-2\sigma} g_{\mu\nu} & 0 & 0 \cr 
0 & e^{2\gamma} e^{-2\sigma} & 0 \cr
0 & 0 & e^{\sigma} \delta_{i\overline j} \cr
}.
\end{equation}
The surviving components of the field strength $G_{MNPQ}$ are only 
$G_{\mu\nu\rho4}$ and $G_{\mu i\overline j4}$, with
\begin{equation}
G_{\mu\nu\rho4} = 3\partial_{[\mu}C_{\nu\rho]4}, ~~
G_{\mu i\overline j4}=\partial_\mu C_{i\overline j4}, ~~
C_{i\overline j4}= ia(x) \, \delta_{i\overline j}. 
\end{equation}
The resulting four-dimensional Lagrangian is
\begin{equation}\label{CJSbulkcomp}
\begin{array}{rcl}
e^{-1}{\cal L}_{\rm CJS} &=& 
-{1\over2\kappa^2}R
-{1\over 4\kappa^2}\left[9(\partial_\mu\sigma)(\partial^\mu\sigma)
+{1\over6}e^{6\sigma} G_{\mu\nu\rho4}G^{\mu\nu\rho4}\right] \crbig
&&-{3\over 4\kappa^2}\left[(\partial_\mu\gamma)(\partial^\mu\gamma)
+e^{-2\gamma} (\partial_\mu a)(\partial^\mu a) \right].
\end{array}
\end{equation}
In this expression, $\kappa$ is the four-dimensional gravitational 
coupling with $\kappa^2 = \kappa_{11}^2/V_7$, $V_7=V_1V_6$ 
being the volume of the compact space $S^1 \times X_6$.

At this stage, the identification of the bosonic components 
$C$, $b_{\mu\nu}$, 
$C_T$ and $T_\mu$ with the bulk fields $\sigma$, $C_{\mu\nu 4}$, 
$\gamma$ and $a$ can only be determined up to two 
proportionality constants (one for each `M-theory multiplet' 
$V$ and $V_T$). These constants can however be 
determined from the couplings of
$C$ and $C_T$ to charged matter and gauge fields \cite{DS1}. The result 
is 
\begin{equation}\label{CJScompident}
\begin{array}{rclrcl}
4\kappa^2 C &=& {\lambda^2\over V_6} e^{-3\sigma}, & \qquad
4\kappa^2 b_{\mu\nu} &=& {\lambda^2\over V_6} C_{\mu\nu 4}, \crbig
C_T &=& 2{\lambda^2\over V_6} e^\gamma, 
& T_\mu &=&  -2{\lambda^2\over V_6} \partial_\mu a.
\end{array}
\end{equation}
The quantity $\lambda$ is the gauge coupling constant on the ${\bf Z}_2$ 
fixed planes. The dimensionless number $\lambda^2/V_6$ actually never 
appears in the four-dimensional effective theory.

\subsubsection{Addition of a superpotential}\label{secsuperpot}

The standard reduction of eleven-dimensional supergravity with
unbroken $N=1$ supersymmetry does not generate a superpotential. This 
fact is however not a direct consequence of the eleven-dimensional 
Bianchi identity or of the CY and $S^1/{\bf Z}_2$ symmetries. In 
principle, the Bianchi identity $\partial_{[M}G_{NPQR]}=0$ allows a 
solution
\begin{equation}\label{Gijk4}
G_{ijk4} = 2i\kappa^{-1}h \epsilon_{ijk}, ~~
G_{\overline {ijk}4} = -2i\kappa^{-1}h \epsilon_{\overline{ijk}},
\end{equation}
where $h$ is a real constant and $\epsilon_{ijk}$ is the
$SU(3)$-invariant CY tensor. The second term in the Lagrangian 
(\ref{11dsugraCJS}) generates then an extra contribution in the 
effective supergravity which corresponds to the addition of a 
superpotential term $[ihS_0^3]_F$ to the bulk Lagrangian. This 
contribution however breaks supersymmetry \cite{DIN1}. Since we 
have insisted in writing Lagrangians in which all Bianchi 
identities are field equations, we prefer to consider
\begin{equation}\label{superpot1}
[U(W+\overline W)]_D + [S_0^3\, W]_F.
\end{equation}
In this way, the fact that the chiral multiplet $W$ ($w=n=0$) 
is an arbitrary imaginary constant is imposed by
the field equation of the vector multiplet $U$ ($w=2$, $n=0$).

With the addition of a superpotential, the bulk Lagrangian takes its 
final `off-shell' form
\begin{equation}\label{LcompCYfinal}
{\cal L}_{\rm B} = \left[ 
-\Upsilon -(S+\overline S)V + L_TV_T
+ U(W+\overline W) \right]_D + [S_0^3 W]_F,
\end{equation}
in which the Bianchi identities of eleven-dimensional supergravity 
are translated into field equations of the Lagrange multipliers $S$, 
$L_T$ and $U$.

\subsection{Modified Bianchi identities and $\kappa$-expansion}
\label{seckappaexp}

Compactification of M-theory on $S^1/{\bf Z}_2$ is usually discussed 
in an expansion in powers of $\kappa_{11}$. Compactification on $O_7$ 
can similarly be formulated with $\kappa$ as expansion parameter. In 
the upstairs version, Bianchi identities are modified at the 
ten-dimensional planes fixed by $S^1/{\bf Z}_2$. Suppose now that 
we modify the four-dimensional supersymmetric Bianchi identities of 
the bulk Lagrangian in the following way (we set $h=0$):
\begin{equation}\label{modif1}
{\cal L}_{\rm B} \quad\longrightarrow\quad {\cal L}=
\left[ -\Upsilon - (S+\overline S)(V + \Delta_V) + L_T(V_T +\Delta_T) 
\right]_D\,,
\end{equation}
with two composite vector multiplets $\Delta_V$ ($w=2$, $n=0$) and 
$\Delta_T$ ($w=n=0$). Solving for the Lagrange multipliers now leads to
$$
V= L-\Delta_V, ~~
V_T=T+\overline T- \Delta_T.
$$
The Lagrangian to first order in these modifications is then
\begin{equation}\label{modif2}
{\cal L} = 
{\cal L}_{\rm B} 
- \left[{\Upsilon\over 2V}\Delta_V 
-{3\over2V_T}(\Upsilon\Delta_T)\right]_D,
\end{equation}
with $V$ and $V_T$ respectively replaced by $L$ and $T+\overline
T$. The multiplets $\Delta_V$ and $\Upsilon\Delta_T$, with
`canonical' dimension $w=2$, appear at order $\Upsilon^0 \sim
\kappa^0$, in comparison with bulk terms of order
$\Upsilon\sim\kappa^{-2}$. This is the relation with the expansion in
powers of $\kappa_{11}$ of M-theory in the low-energy limit. In
M-theory compactification, the multiplets $\Delta_V$ and $\Delta_T$
can thus be obtained either by considering the modified Bianchi 
identities on $O_7$, formulated as in Eq. (\ref{modif1}), or from 
corrections to the Lagrangian of eleven-dimensional supergravity on 
$O_7$, as in expression (\ref{modif2}).

\section{Gauge and matter contributions from the two ${\bf Z}_2$ 
fixed planes}\label{secplanes}
\setcounter{equation}{0}

In this section, we show that the introduction of the next to lowest 
order corrections (gauge multiplets and charged matter contributions) 
is controlled by a simple modification of the four-dimensional Bianchi 
identities, in analogy with the appearance of ${\bf Z}_2$ fixed planes 
contributions in the M-theory Bianchi identities. 

We start by considering the well-known dependence on charged matter
(in chiral multiplets collectively denoted by $M$, with $w=n=0$) and 
gauge multiplets (vector multiplet $A$, in the adjoint representation, 
with $w=n=0$) of the effective $N=1$ four-dimensional supergravity for 
CY compactifications of heterotic strings 
\cite{W4dsugra, DIN, BFQ}. The Lagrangian in the chiral
formulation (\ref{bulkchiral}) becomes
\begin{equation}\label{matter1}
{\cal L}_{\rm c} = 
-{3\over2}\left[S_0\overline S_0 e^{-{K/ 3}}\right]_D
+ \left[S_0^3W\right]_F
+{1\over4}\left[S{\cal WW}\right]_F, 
\end{equation}
with
\begin{equation}
K = -\log(S+\overline S)-3\log(T+\overline T - 2\overline Me^AM) 
\end{equation}
and $W = \alpha M^3$. 
The superpotential should be understood as a gauge invariant trilinear 
interaction with coupling constant $\alpha$ defined as an integral
over the CY space.
The chiral multiplet ${\cal W}$ ($w=n=3/2$) is the gauge
field-strength for $A$. 
The gauge group is in general not simple, and
\begin{equation}\label{Wdef}
{\cal WW} = \sum_a c^a{\cal W}^a{\cal W}^a,
\end{equation}
with a real coefficient $c^a$ for each simple or abelian factor.
In the linear equivalent version of the theory, the Lagrangian 
(\ref{4dLCJS}) reads now \cite{CFV, DQQ}
\begin{equation}\label{matter2}
{\cal L}_{\rm l} = 
- \left[ (S_0\overline S_0e^{-\hat K/3})^{3/2}(2\hat L)^{-1/2} 
\right]_D
+\, [\alpha S_0^3 M^3]_F,
\end{equation}
where the new modulus and matter K\"ahler potential is
\begin{equation}\label{hatKis2}
\hat K = -3\log(T+\overline T - 2\overline M e^A M).
\end{equation}
The linear multiplet $L$ is replaced by
\begin{equation}\label{Lhatis}
\hat L = L - 2\Omega,
\end{equation}
with the Chern-Simons vector multiplet $\Omega$ ($w = 2$, $n=0$) 
defined by\footnote{
In global Poincar\'e supersymmetry, $\Sigma(\Omega) = 
-{1\over4}\overline{DD} \Omega$.} 
\begin{equation}\label{CSdef}
\Omega = \sum_a c^a\Omega^a, ~~
\Sigma(\Omega^a) = {1\over16}{\cal W}^a{\cal W}^a.
\end{equation}

Insisting as before on Bianchi identities, both forms 
(\ref{matter1}) and (\ref{matter2}) are equivalent to 
\begin{equation}\label{matter3}
\begin{array}{rcl}
{\cal L} &=&
\left[ -\Upsilon -(S+\overline S)(V+2\Omega)
+ L_T(V_T+2\overline Me^AM) \right]_D \crbig 
&&+\left[U(W-\alpha M^3)+ {\rm c.c.}\right]_D
+ \left[S_0^3 W\right]_F \crbig 
&=& 
\left[ -\Upsilon -(S+\overline S)(V+2\Omega)
+ L_T(V_T+2\overline Me^AM) \right]_D \crbig
&&+ \left[S_0^3(ih +\alpha M^3)\right]_F.
\end{array}
\end{equation}
Supersymmetric vacua have $h=0$.
As before, solving for $S$ and $L_T$ imposes respectively 
$V=L-2\Omega=\hat L$ and $V_T=T+\overline T-2\overline Me^AM$, 
leading to Eq. (\ref{matter2}). Alternatively, with the tensor 
calculus identity (and up to an irrelevant total derivative)
\begin{equation}\label{DFident}
-2[(S+\overline S)\Omega]_D = 
{1\over4}\sum_ac^a[S{\cal W}^a {\cal W}^a]_F,
\end{equation}
the resolution for $V$ and $L_T$ leads back to the chiral form 
(\ref{matter1}).

This reformulation of the gauge invariant Lagrangian suggests some 
remarks. 
First\-ly, it enhances the importance of gauge {\it and} matter 
Chern-Simons multiplets in superstring effective actions.
Secondly, the Chern-Simons vector multiplet $\Omega(A)$ is not gauge 
invariant: its variation is a linear multiplet. The variation of
$[(S+\overline S)2\Omega]_D$ is then a derivative and $V$ remains gauge 
invariant. When solving for $S$, it simply follows that $\hat L$ is 
gauge invariant and that the linear multiplet transforms as 
$\delta L = 2\delta\Omega$.
Finally, expression (\ref{matter3}) shows that all gauge and 
chiral matter contributions can be viewed as the supersymmetrization 
of modified Bianchi identities imposed by $S$, $L_T$ and $U$. 
This observation provides the link to the approach based on M-theory 
on $O_7$, in which the ${\bf Z}_2$ fixed planes carrying the 
Yang-Mills fields induce because of supersymmetry modifications 
to the Bianchi identity of the four-form field strength of 
eleven-dimensional supergravity.

In the effective supergravity of M-theory on $O_7$ (`upstairs 
formulation'), the various components of the Lagrangian 
(\ref{matter3}) have the following origin. 
The first term is the bulk supergravity contribution. 
The second term, $[(S+\overline S)(V+2\Omega)]_D$, is the 
supersymmetrization 
of the Bianchi identity verified by the component $G_{\mu\nu\rho4}$ 
of the field $G$, modified by gauge contributions on the fixed
planes. Similarly, the two last terms, $[L_T(V_T+2\overline Me^AM)]_D$ and 
$[U(W-\alpha M^3)+{\rm c.c.}]_D$, are respectively the supersymmetric 
extensions of the Bianchi identities of $G_{\mu j\overline k 4}$ and 
$G_{ijk4}$. All the fixed plane contributions are then given at this order 
by the supersymmetrization of Bianchi identities, as obtained by
direct $O_7$ truncation of the eleven-dimensional identities 
\cite{HW1, HW2}. 

At this point, the gauge coupling constant for each simple or 
abelian factor $a$ in the gauge group appears to be
\begin{equation}\label{gis1}
{1\over g_a^2} = c^a \Re s.
\end{equation}
At this order, $g_a$ is the tree-level wilsonnian and 
physical\footnote{The coefficient of 
$-{1\over4}F_{\mu\nu}^aF^{a\mu\nu}$ in the generating functional 
of one-particle irreducible Green's functions.} gauge coupling.

It is clear, as already observed \cite{ref1}--\cite{LOW1}, that as far 
as the structure of the four-dimensional effective supergravity is 
concerned, the same information follows from $O_7$ compactification 
of M-theory at the next to lowest order in the $\kappa$-expansion and 
from CY compactifications of the heterotic strings, at zero string 
loop order.

\section{Anomaly-cancelling terms}\label{secanom}
\setcounter{equation}{0}

In the ten-dimensional heterotic string, cancellation of gauge and 
gravitational anomalies is a one-loop effect in string or effective 
supergravity perturbation theory. In four space-time dimensions, the 
nature of the cancelled anomalies is known from studies of $(2,2)$ 
compactifications of heterotic strings in the Yang-Mills sector 
\cite{DKL, DFKZ1, L}: target-space duality of the modulus $T$ has a 
one-loop anomaly which is cancelled by a counterterm in the one-loop 
Wilson Lagrangian ${\cal L}^{(1)}_W$\footnote{
The expressions given in the previous sections were for 
${\cal L}^{(0)}_W$, or for the tree-level standard effective 
Lagrangian ${\cal L}_\Gamma$.}, in a generalization to sigma-model 
anomalies of the Green-Schwarz mechanism \cite{GSsigma}. The
derivation of the complete counterterm requires a calculation to all 
orders in the modulus $T$ \cite{DKL}. However, at the present stage 
of understanding, the M-theory approach should be regarded as a 
large-$T$ limit in which T-duality reduces to a shift symmetry in 
the imaginary part of $T$.

In the large-$T$ limit, the $T$-dependent corrections to gauge kinetic 
terms are of the form (see ref. \cite{DS1} and citations therein)
\begin{equation}\label{gaugthr1}
{1\over4}\sum_a \beta^a \left[ T{\cal W}^a{\cal W}^a\right]_F,
\end{equation}
where the coefficients $\beta^a$ are in principle calculable in 
heterotic strings. Taking also into account the $D$-density 
$\left[L_T(V_T + 2\overline Me^AM)\right]_D$ present in 
Lagrangian (\ref{matter3}), we can rewrite expression 
(\ref{gaugthr1}) in terms of the `M-theory multiplets':
\begin{equation}\label{gaugthr3}
\Bigl[(L_T-2\sum_a\beta^a\Omega^a)(V_T+2\overline Me^AM)\Bigr]_D.
\end{equation}

The correction (\ref{gaugthr1}) to the SYM Lagrangian is independent 
of the matter fields and can be seen as a correction to the
holomorphic gauge kinetic function $f_{ab}$. A possible 
matter-dependent contribution to gauge kinetic terms is the 
gauge invariant real density\footnote{
For simplicity, we consider the standard embedding with a gauge 
group $E_6 \times E_8$, with the notation 
$\Omega = \Omega^1 +\Omega^2$, and with a matter multiplet $M$ 
transforming as ({\bf 27},{\bf 1}) of $E_6\times E_8$.
}:
\begin{equation}
-2\delta\Bigl[\overline M e^A M(L-2\sum_{a=1}^{2}{\Omega^a})\Bigr]_D,
\end{equation}
or
\begin{equation}
-2\delta\Bigl[\overline M e^A M\, V\Bigr]_D,
\end{equation}
using the `M-theory multiplet' $V$.

The M-theory anomaly-cancelling terms generate a further contribution 
of the form \cite{DS1}
\begin{equation}\label{GSsuperpot}
\epsilon \Bigl[ V |\alpha M^3|^2 \Bigr]_D.
\end{equation}
In summary, the Wilson Lagrangian up to string one-loop order is 
expected to become
\begin{equation}\label{matter4}
\begin{array}{rcl}
{\cal L} &=& \left[-\Upsilon-(S+\overline S)(V+2\Omega)
+(L_T-2\sum_{a=1}^{2}{\beta^a\Omega^a})
(V_T+2\overline Me^AM)\right]_D \crbig
&&+\left[U(W-\alpha M^3) +{\rm c.c.}\right]_D 
+ \left[S_0^3W\right]_F \crbig
&&+ \left[V(\epsilon |\alpha M^3|^2 -2\delta\overline Me^AM)
\right]_D.
\end{array}
\end{equation}
Each of the one-loop corrections, with coefficients $\beta^1$, 
$\beta^2$, $\epsilon$ and $\delta$ is related to a well-defined 
counterterm which can be easily identified in the KK reduction of 
the ten-dimensional Green-Schwarz counterterms arising from M-theory 
on $S^1/{\bf Z}_2$ \cite{HW1, HW2, GX7}. An explicit computation 
predicts in particular the relations 
$\beta^1=-\beta^2=\delta$ \cite{DS1}.

From the general expression (\ref{matter4}), we can derive various 
equivalent forms. For instance, solving for $S$, $L_T$ and $U$ gives 
the version of the effective supergravity in which the dilaton is 
described by a linear multiplet:
\begin{equation}\label{linearvers}
\begin{array}{rcl}
{\cal L}_{\rm l} &=&
\left[-(S_0\overline S_0)^{3/2}\left(T+\overline T-2 
\overline Me^AM\right)^{3/2}
(2\hat L)^{-{1/2}} \right]_D 
+\left[S_0^3(ih +\alpha M^3)\right]_F \crbig
&&+\,{1\over 4}\left[T\sum_{a=1}^{2}{\beta^a{\cal W}^a{\cal W}^a}
\right]_F
+\left[\hat L \left(\epsilon|\alpha M^3|^2-2\delta\overline Me^AM\right)
\right]_D.
\end{array}
\end{equation}
The threshold corrections are the holomorphic $T$-dependent terms 
controlled by $\beta^1$ and $\beta^2$.

We can also solve for $V$, $L_T$ and $U$ in Eq. (\ref{matter4})
to get the version with a chiral dilaton multiplet:
\begin{equation}\label{chiralvers}
{\cal L}_{\rm c} = 
-{3\over 2}\left[S_0\overline S_0\,e^{-K/3}\right]_D
+\left[S_0^3(ih +\alpha M^3)\right]_F
+{1\over 4}
\sum_{a=1}^{2}{
\left[
(S+\beta^aT){\cal W}^a{\cal W}^a\right]_F},
\end{equation}
with the K\"ahler potential
\begin{equation}\label{fullK1}
K = -\log\left(S+\overline S+2\delta\overline Me^AM
-\epsilon|\alpha M^3|^2\right) 
-3\log\left(T+\overline T-2\overline Me^AM\right),
\end{equation}
and the gauge kinetic functions $f^a=S+\beta^aT$. 
The term with coefficient $\delta$ has been obtained in direct CY 
reductions of M-theory on $S^1/{\bf Z}_2$ (see for instance 
\cite{NOY,LOW1}). The charged matter contribution with coefficient 
$\epsilon$ was not included in these analyses. 
Observe however that an ambiguity exists because of the possibility 
to perform a holomorphic redefinition of the two chiral multiplets 
$S$ and $T$. To remove this ambiguity, we can use information from 
M-theory compactification \cite{DS1}, or choose the 
unequivocal linear version.

The gauge contributions appearing in Eq. (\ref{matter4}) read
$$
-2\sum_{a=1}^{2}\left[\left
(S+\overline S+\beta^a(V_T+2\overline Me^AM)\right)\Omega^a\right]_D,
$$
so that the gauge coupling constants are given by
\begin{equation}
{1\over g_a^2}=\Re s+{1\over 2}\beta^a\left(C_T+2
\overline MM \right).
\end{equation}
This expression becomes harmonic once the Bianchi identity 
imposing $C_T+2\overline MM=2\Re T$ has been used.

\newpage
\vspace{.5cm}
\begin{center}{\bf Acknowledgments}\end{center}
\noindent
This research has been supported in part by
the European Union under the TMR contract ERBFMRX-CT96-0045,
the Swiss National Science Foundation
and the Swiss Office for Education and Science.

\end{document}